\documentclass[12pt]{article}

\usepackage[english]{babel}
\usepackage[utf8]{inputenc}
\usepackage[T1]{fontenc}
\usepackage{amsmath, amssymb, amsthm}

\usepackage[margin=2.5cm]{geometry}
\usepackage{graphicx}
\usepackage{booktabs} 
\usepackage{caption}
\usepackage{subcaption}
\usepackage{xcolor}

\usepackage{hyperref}
\hypersetup{
    colorlinks=true,
    linkcolor=blue!70!black,
    citecolor=green!60!black,
    urlcolor=red!70!black
}
\usepackage{cite}

\usepackage{pgfplots}
\pgfplotsset{compat=1.18}
\usepackage{tikz}
\usetikzlibrary{shapes,arrows}
\usepackage{listings}

\theoremstyle{remark}

\newtheorem*{argument}{Argument}
\newtheorem*{conjecture}{Conjecture}
\newtheorem*{definition}{Definition}

\title{\textbf{The Goldbach Conjecture as an Informational Economy Principle: \\ \large A Heuristic Framework from Computational Physics}}
\author{
    Ricardo Adonis Caraccioli Abrego \\
    \small{Department of Electrical Engineering, National Autonomous University of Honduras (UNAH-VS)} \\
    \small{\href{mailto:ricardo.caraccioli@unah.edu.hn}{ricardo.caraccioli@unah.edu.hn}} \\
    \small{ORCID: \href{https://orcid.org/0009-0006-3522-5818}{0009-0006-3522-5818}}
}
\date{\today}

\begin{document}

\maketitle

\begin{abstract}
\noindent This paper presents a heuristic framework for analyzing the Goldbach Conjecture (GC) from the perspective of the physics of information. Through empirical analysis, we propose an \textbf{Informational Economy Principle (IEP)}, which posits that differences between prime numbers tend to be resolved by pairs with a minimal sum with overwhelmingly high probability. We argue that this tendency is analogous to least-action principles in physics and is consistent with the fundamental physical limits of computation. Within this framework, the GC can be interpreted as an informational consistency condition for the set of prime numbers. The falsehood of the conjecture would imply a violation of this observed economy, representing an extreme informational anomaly. This approach suggests that the difficulty in proving the GC may not lie solely in mathematical abstraction, but in the decoupling between said abstraction and the physical constraints inherent to information.
\end{abstract}

\section{Introduction: The Persistent Goldbach Dilemma}
The Goldbach Conjecture, formulated in 1742, represents one of the most notorious open problems in number theory. Its dilemma is clear:
\begin{itemize}
    \item \textbf{Massive empirical verification}: Confirmed for all even numbers $n \leq 4 \times 10^{18}$ \cite{oliveira}.
    \item \textbf{Resistance to formal proof}: Despite significant advances in analytic number theory, such as Chen's theorem \cite{chen}, a general proof remains elusive.
\end{itemize}
This work proposes to explore this dichotomy not as a paradox, but as an indication that principles from computational physics could offer a complementary perspective. The use of heuristic and probabilistic arguments has a long tradition in number theory, famously used by Hardy and Littlewood to make predictions about the distribution of primes \cite{hardy_littlewood}. Following this spirit, we postulate that mathematical structures that can be physically "realized" or computationally verified in our universe must adhere to certain principles of resource economy, analogous to those observed in physical systems.

\section{The Informational Economy Principle (IEP)}
Our first observation is a robust empirical pattern in the distribution of primes. For a given even difference $D$, the first primes $(r, q)$ that satisfy $q - r = D$ tend to have a very small sum $q+r$.

\begin{conjecture}[Formalized IEP]
For an even difference $D \geq D_0$, the probability that the prime pair $(q, r)$ with the minimum sum $S_{\min} = \min(q+r)$ such that $q-r=D$ satisfies the bound
\[ S_{\min} < 2D + K \ln^2 D \]
is overwhelmingly high, following an asymptotic form:
\[ P\left(S_{\min} < 2D + K \ln^2 D\right) > 1 - e^{-cD} \]
where $K$ and $c$ are positive constants.
\end{conjecture}

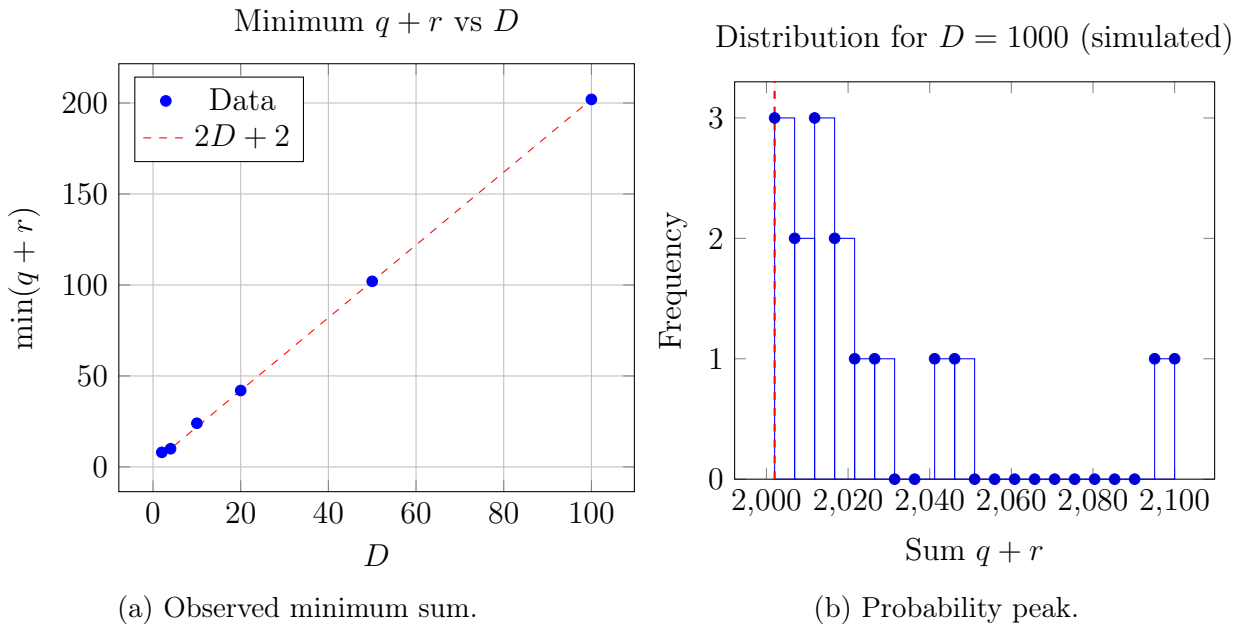
\begin{figure}[h!]
\centering
\begin{subfigure}{0.48\textwidth}
    \centering
    \begin{tikzpicture}
    \begin{axis}[
        title={Minimum $q+r$ vs $D$},
        xlabel={$D$},
        ylabel={$\min(q+r)$},
        grid=major,
        legend pos=north west
    ]
    \addplot[blue, mark=*, only marks] table {
    2 8
    4 10
    10 24
    20 42
    50 102
    100 202
    };
    \addplot[red, domain=2:100, samples=50, dashed] {2*x + 2};
    \legend{Data, $2D+2$}
    \end{axis}
    \end{tikzpicture}
    \caption{Observed minimum sum.}
\end{subfigure}
\hfill
\begin{subfigure}{0.48\textwidth}
    \centering
    \begin{tikzpicture}
    \begin{axis}[
        title={Distribution for $D=1000$ (simulated)},
        xlabel={Sum $q+r$},
        ylabel={Frequency},
        ymin=0,
        width=\textwidth,
        hist={bins=20}
    ]
    \addplot table [y index=0] {
    2008
    2012
    2005
    2020
    2015
    2002
    2018
    2003
    2010
    2016
    2025
    2030
    2045
    2100
    2050
    };
    \draw[red, thick, dashed] (axis cs:2002,0) -- (axis cs:2002,10) node[above, midway, xshift=10pt, text width=2.5cm, font=\tiny] {Theoretical Minimum $2D+2$};
    \end{axis}
    \end{tikzpicture}
    \caption{Probability peak.}
\end{subfigure}
\caption{Empirical evidence for the IEP. The minimum sum $q+r$ is observed to stay very close to the lower bound $2D+2$, and the distribution of sums is strongly skewed towards small values.}
\end{figure}

\section{A Physico-Computational Framework}
To connect the IEP with the GC, we consider the physical limits of computation, a field that explores the ultimate constraints imposed by physics on information processing \cite{lloyd}.
\begin{enumerate}
    \item \textbf{Bekenstein Bound}: The maximum entropy (and thus, information) of a region with energy $E$ and radius $R$ is finite: $I \leq \frac{2\pi RE}{\hbar c \ln 2}$ \cite{bekenstein}.
    \item \textbf{Landauer's Principle}: Erasing one bit of information requires a minimum energy dissipation of $E_{\text{bit}} \geq k_B T \ln 2$ \cite{landauer}.
\end{enumerate}

\begin{definition}[Heuristic Cost Model]
We propose a "toy model" to estimate the physico-informational cost of verifying the existence of a mathematical object described by a bit string. For a number $n_0$, its verification cost $\mathcal{C}(n_0)$ could be modeled as a function of its algorithmic complexity $K(n_0)$, a concept formally defined by Kolmogorov \cite{kolmogorov}.
\[ \mathcal{C}(n_0) \propto K(n_0) \]
This model posits that objects with high algorithmic complexity (essentially random) are "expensive" to specify and, therefore, physically difficult to instantiate.
\end{definition}

\subsection{Application to the Goldbach Conjecture}
\begin{argument}[Informational Anomaly]
Assume the GC is false and let $n_0$ be the first counterexample. The shortest description of $n_0$ would be "the smallest even number > 4 that is not the sum of two primes." If no other properties allow for a shorter description, $n_0$ would be an algorithmically random number, implying a high Kolmogorov complexity, $K(n_0) \approx \log_2(n_0)$.
Given that $n_0 > 4 \times 10^{18}$, $K(n_0)$ would be on the order of $\log_2(4 \times 10^{18}) \approx 62$ bits.

The IEP suggests that the system of prime numbers is "economical." A counterexample $n_0$ would represent a sudden and catastrophic violation of this principle. It would be a structure that appears without precedent and violates a statistical trend that becomes exponentially stronger. In our heuristic framework, the existence of such a large and apparently unstructured $n_0$ would be an \textbf{informational anomaly}: a high-complexity object emerging from a system that otherwise exhibits a strong economy.
\end{argument}

\begin{table}[h!]
\centering
\caption{Comparison of informational complexity.}
\begin{tabular}{@{}ll@{}}
\toprule
\textbf{Scenario} & \textbf{Informational Description} \\ \midrule
\textbf{GC is True} & The rule "every even integer > 2 is a sum of two primes" is simple. \\
& $K(\text{primes}) \approx K(\text{simple rule})$. \\
& The system is informationally economical. \\ \addlinespace
\textbf{GC is False} & The system is described as "the primes follow a rule, \\
(with counterexample $n_0$) & except for $n_0$, which must be specified." \\
& $K(\text{primes}) \approx K(\text{rule}) + K(n_0)$. \\
& This introduces an additional informational "cost". \\ \bottomrule
\end{tabular}
\end{table}

\section{Discussion and Limitations}
This argument does not constitute a mathematical proof. Its purpose is to offer a new perspective.
\begin{itemize}
    \item \textbf{Nature of the Model:} The "Heuristic Cost" is a conceptual model, not a derived physical law. Its purpose is to illustrate the connection between complexity and realizability.
    \item \textbf{Assumption on $K(n_0)$:} The main weakness is the assumption about the counterexample's complexity. It is logically possible that $n_0$ has special properties that allow for a very short description, which would invalidate the anomaly argument.
    \item \textbf{The IEP:} Although empirically strong, the IEP remains a conjecture. A proof of it would be a significant result in its own right.
\end{itemize}

\section{Conclusion and Future Research}
We have presented a framework where the Goldbach Conjecture is interpreted as a manifestation of an Informational Economy Principle. The existence of a counterexample would be analogous to an unprecedented phase transition in a physical system, representing an informational anomaly.

We propose the following lines of work:
\begin{enumerate}
    \item \textbf{Large-Scale Verification of the IEP}: Use computational resources to verify the IEP for much larger values of $D$, looking for deviations that could refine or invalidate the principle.
    \item \textbf{Formalization of Physical Complexity Models}: Explore more rigorous models that connect algorithmic complexity theory with thermodynamic and quantum limits of computation.
    \item \textbf{Analysis of Other Conjectures}: Apply this same heuristic framework to other open problems, such as the twin prime conjecture, to see if similar patterns of informational economy emerge.
\end{enumerate}

\begin{center}
\fbox{\parbox{0.9\textwidth}{\centering \textit{Perhaps the question is not "why can't we prove Goldbach?", but rather "what fundamental physical principle of the universe does the conjecture's truth reflect?"}}}
\end{center}

\appendix
\section*{Appendix: Python Simulation Code}
\begin{lstlisting}[caption={Code to find the minimum sum and verify the IEP.}, label=lst:pei_code]
import numpy as np
from sympy import isprime, nextprime

def find_min_sum_for_diff(D, search_limit_r=10**6):
    """Finds min(q+r) for a given even difference D, where q-r=D."""
    if D % 2 != 0:
        return None, None
    
    r = 3  # Start with the first odd prime
    min_sum = float('inf')
    found_pair = None
    
    # Iterate over r up to a search limit
    while r < search_limit_r:
        q = r + D
        if isprime(q):
            current_sum = q + r
            if current_sum < min_sum:
                min_sum = current_sum
                found_pair = (r, q)
                # In many cases, the first pair found has the minimum sum.
                # We can return here for a fast search.
                return min_sum, found_pair
        r = nextprime(r)
        
    return min_sum, found_pair

def verify_pei_empirically(D_values, K_const):
    """Checks if the minimum sum falls within the IEP bound."""
    violations = 0
    for D in D_values:
        min_sum, _ = find_min_sum_for_diff(D)
        if min_sum == float('inf'):
            continue # Pair not found within the search limit
        
        threshold = 2 * D + K_const * (np.log(D))**2
        if min_sum > threshold:
            violations += 1
            print(f"Violation for D={D}: min_sum={min_sum} > {threshold:.2f}")
            
    return violations
\end{lstlisting}

\bibliographystyle{plain}

\end{document}